\documentclass[twocolumn,showpacs,superscriptaddress,amsmath,amssymb]{revtex4}
\usepackage{amsmath}
\usepackage{float}
\usepackage{graphicx}
\usepackage{dcolumn}
\begin{document}

\title{Ground-state properties of closed-shell nuclei with low-momentum
realistic interactions}
\author{L. Coraggio}
\affiliation{Dipartimento di Scienze Fisiche, Universit\`a
di Napoli Federico II, \\ and Istituto Nazionale di Fisica Nucleare, \\
Complesso Universitario di Monte  S. Angelo, Via Cintia - I-80126 Napoli,
Italy}
\author{N. Itaco}
\affiliation{Dipartimento di Scienze Fisiche, Universit\`a
di Napoli Federico II, \\ and Istituto Nazionale di Fisica Nucleare, \\
Complesso Universitario di Monte  S. Angelo, Via Cintia - I-80126 Napoli,
Italy}
\author{A. Covello}
\affiliation{Dipartimento di Scienze Fisiche, Universit\`a
di Napoli Federico II, \\ and Istituto Nazionale di Fisica Nucleare, \\
Complesso Universitario di Monte  S. Angelo, Via Cintia - I-80126 Napoli,
Italy}
\author{A. Gargano}
\affiliation{Dipartimento di Scienze Fisiche, Universit\`a
di Napoli Federico II, \\ and Istituto Nazionale di Fisica Nucleare, \\
Complesso Universitario di Monte  S. Angelo, Via Cintia - I-80126 Napoli,
Italy}
\author{T. T. S. Kuo}
\affiliation{Department of Physics, SUNY, Stony Brook, New York 11794}

\date{\today}

\begin{abstract}
Ground-state properties of $^{16}$O and $^{40}$Ca are calculated with 
a low-momentum nucleon-nucleon potential, $V_{\rm low-k}$, derived from 
the chiral N$^3$LO interaction recently constructed by Entem and Machleidt. 
The smooth $V_{\rm low-k}$ is used directly in a Hartree-Fock approach, 
avoiding the difficulties of the Brueckner-Hartree-Fock procedure. 
Corrections up to third order in the Goldstone expansion are evaluated, 
leading to results that are in very good agreement with experiment.
Convergence properties of the expansion are examined.
\end{abstract}


\maketitle

\section{Introduction}

A fundamental problem in nuclear theory has long been the calculation 
of the bulk properties of closed--shell nuclei, such as their binding 
energy and charge radius, starting from realistic nucleon--nucleon ($NN$) 
potentials.
Potentials like CD--Bonn \cite{cdbonn}, Nijmegen \cite{stoks94}, 
Argonne $v_{18}$ \cite{wiringa95}, and the new chiral potential of 
Ref. \cite{mach03} reproduce the $NN$ scattering data and the observed 
deuteron properties very accurately, but, because of their strong 
repulsion at short distances, none of them can be used directly in 
nuclear structure calculations.

A traditional approach to this problem is the Brueckner-Goldstone (BG)
theory \cite{Day67}, where the Goldstone perturbative expansion is 
re-ordered summing to all orders a selected class of diagrams,
the ladder diagrams. 
This implies replacing the bare interaction ($V_{NN}$) vertices by the 
reaction matrix ($G$) ones and omitting the ladder diagrams.
Within this framework, one has the well known Brueckner--Hartree--Fock 
(BHF) theory when the self--consistent definition is adopted for the
single--particle (SP) auxiliary potential and only the first--order
contribution in the BG expansion is taken into account. 
The BHF approximation gives therefore a mean field description of the 
ground state of nuclei in terms of the $G$ matrix, the latter taking 
into account the correlations between pairs of nucleons. 
However, owing to the energy dependence of $G$, this procedure is not 
without difficulties. 
An important issue is the choice of the SP energies for states above 
the Fermi surface, which  are not uniquely defined \cite{towner}.

Calculations for finite nuclei within the BHF approach lead usually to
insufficient binding energy as well as too small charge radii 
\cite{sch91}.
In this context, extensions of the conventional BHF approach have been
proposed to account for long--range correlations \cite{gad02}.
It is worth mentioning that alternative approaches have been developed, 
as for instance the correlated basis function method \cite{fabrocini00} 
and the coupled cluster method \cite{heisenberg99}, in both of which 
correlations are directly embedded into the wave functions.
A comprehensive review of the various methods is given in Ref.
\cite{muther00}, which also includes a discussion of  calculations for
nuclear matter as well as for finite nuclei. 
An other effort in this direction is the unitary model--operator 
approach of Suzuki and Okamoto \cite{suzuki94}.

Recently, a new technique to renormalize the short--range repulsion of 
a realistic $NN$ potential by integrating out its high momentum 
components has been proposed \cite{bogner01,bogner02}. 
The resulting low--momentum potential, which we call $V_{\rm low-k}$, 
is a smooth potential that preserves the low--energy physics of $V_{NN}$ 
and is therefore suitable for being used directly in nuclear structure 
calculations.  
We have employed $V_{\rm low-k}$ derived from modern $NN$ potentials to 
calculate shell--model effective interactions by means of the 
$\hat{Q}$--box plus folded diagram method. 
Within this framework, several nuclei with few valence particles 
have been studied \cite{covello02,coraggio02,coraggio02a}, leading to the 
conclusion that $V_{\rm low-k}$ is a valid input for realistic shell-model 
calculations.

The main purpose of this paper is to show that this potential may also 
be profitably used for the calculation of ground state properties of 
doubly closed nuclei. 
To this end, we have employed the Goldstone expansion  which, given the 
smooth  behavior of $V_{\rm low-k}$, does not require any rearrangement. 
As a first step of our procedure, we solve the Hartree--Fock (HF) 
equations for $V_{\rm low-k}$. 
Then, using the obtained self--consistent field as auxiliary potential, 
we calculate the Goldstone expansion including diagrams up to third order 
in $V_{\rm low-k}$.

Here, we present the results obtained for $^{16}$O and $^{40}$Ca starting
from the new realistic $NN$ potential of Entem and Machleidt \cite{mach03}
based on chiral perturbation theory at the next-to-next-to-next-to-leading
order (N$^3$LO; fourth order). This potential is an improved version of 
the earlier chiral $NN$ potential \cite{mach02} (known as Idaho potential)
constructed by the same authors, which includes two-pion exchange
contributions only up to chiral order three.
We have recently employed the latter potential in shell-model calculations
for various two-particle valence nuclei \cite{coraggio02}.

The paper is organized as follows. 
In Sec. II we first describe the main features of the derivation of 
$V_{\rm low-k}$, then outline the essentials of our calculation. 
In Sec. III we present our results and compare them with the experimental 
data. 
Some concluding remarks are given in Sec. IV.

\section{Method of calculation}
The first step in our approach is to integrate out the high-momentum
components of $V_{NN}$. 
According to the general definition of a renormalization group
transformation, the decimation must be such that the low-energy 
observables calculated in the full theory are preserved exactly by the 
effective theory.
Once the relevant low-energy modes are identified, all remaining modes 
or states have to be integrated out.

For the nucleon-nucleon problem in vacuum, we require that the deuteron
binding energy, low-energy phase shifts, and low-momentum
half-on-shell $T$ matrix calculated from $V_{NN}$ must be reproduced
by $V_{\rm low-k}$.

The full-space Schr\"odinger equation may be written as
\begin{equation}
H \Psi_{\mu} = E_{\mu} \Psi_{\mu};~ H=H_0 + V_{NN},
\end{equation}
where $H_0$ is the unperturbed Hamiltonian, namely, the kinetic energy.
The above equation can be reduced to a model-space one of the form
\begin{equation}
PH_{\rm eff}P \Psi_{\mu} = E_{\mu} P \Psi_{\mu};~ H_{\rm eff}=H_0 +
V_{\rm low-k},
\end{equation}
where $P$ denotes the model-space, which is defined by momentum
$k\leq k_{\rm cut}=\Lambda$, $k$ being the relative momentum and $k_{\rm
cut}$
a cut-off momentum.

The half-on-shell $T$ matrix of $V_{NN}$ is

\[ T(k',k,k^2) = V_{NN}(k',k) + ~~~~~~~~~~~~~~~~~~~~~~~~\]
\begin{equation}
~~+\int _0 ^{\infty} q^2 dq  V_{NN}(k',q)
\frac{1}{k^2-q^2 +i0^+ } T(q,k,k^2 ) ~~,
\end{equation}

\noindent
and the effective low-momentum T matrix is defined by

\[T_{\rm low-k }(p',p,p^2) = V_{\rm low-k }(p',p) + ~~~~~~~~~~~~~~~~~~~~~~\]

\begin{equation}
\int _0 ^{\Lambda} q^2 dq
V_{\rm low-k }(p',q) \frac{1}{p^2-q^2 +i0^+ } T_{\rm low-k} (q,p,p^2)~~.
\end{equation}

\noindent
Note that for $T_{\rm low-k }$ the intermediate states are integrated up
to $\Lambda$.

It is required that, for $p$ and $p'$ both belonging to $P$ ($p,p' \leq
\Lambda$), $T(p',p,p^2)= T_{\rm low-k }(p',p,p^2)$.
In Refs. \cite{bogner01,bogner02} it has been shown that the above
requirements are satisfied when $V_{\rm low-k}$ is given by the
folded-diagram
series

\begin{equation}
V_{\rm low-k} = \hat{Q} - \hat{Q'} \int \hat{Q} + \hat{Q'} \int \hat{Q}
\int \hat{Q} - \hat{Q'} \int \hat{Q} \int \hat{Q} \int \hat{Q} + ...~~,
\end{equation}

\noindent
where $\hat{Q}$ is an irreducible vertex function, in the sense that
its intermediate states must be outside the model space $P$.
The integral sign represents a generalized folding operation \cite{krenc80},
and $\hat{Q'}$ is obtained from $\hat{Q}$ by removing terms of first
order in the interaction.

The above $V_{\rm low-k}$ can be calculated by means of iterative
techniques.
We have used here an iteration method proposed in Ref. \cite{andre96}, 
which is particularly suitable for non-degenerate model spaces.
This method, which we refer to as Andreozzi-Lee-Suzuki (ALS) method, 
is an iterative method of the Lee-Suzuki type, which converges to the 
lowest $d$ eigenvalues of $H$, $d$ being the dimension of the $P$ space.
Since the $V_{\rm low-k}$ obtained by this technique is non-hermitian, 
we have made use of the simple and numerically convenient hermitization
procedure suggested in Ref. \cite{andre96}. 
We have verified that the deuteron binding energy and the phase shifts 
up to the cut-off momentum $\Lambda$ are preserved by $V_{\rm low-k}$.

An important question in our approach is what value one should use for 
the cut-off momentum. 
A discussion of this point as well as a criterion for the choice
of $\Lambda$ can be found in Ref. \cite{bogner02}. 
According to this criterion, we have used here $\Lambda=2.1 \; 
{\rm fm}^{-1}$.

\begin{figure}[H]
\includegraphics[scale=1.0,angle=+90]{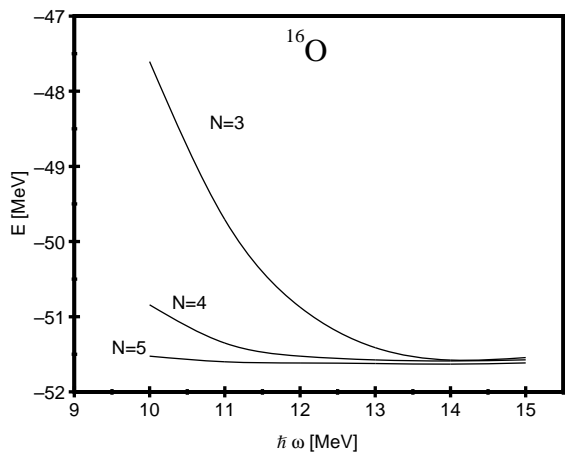}
\caption{Behavior of $E_{\rm HF}$ with $\hbar \omega$ and $N$ for 
$^{16}$O.}
\end{figure}

We have found it appropriate, however, to check on the sensitivity of
our results to moderate changes in the value of $\Lambda$.
It has turned out that they change very little when letting $\Lambda$ 
vary from $2.0$ to $2.2 \; {\rm fm}^{-1}$.
For instance, at the second order in the Goldstone expansion (see next
section), the binding energy per nucleon for $^{16}$O is 7.29 and 
7.12 MeV for $\Lambda=2.0$ and $\Lambda=2.2 \; {\rm fm}^{-1}$, 
respectively.

As already mentioned in the Introduction, our starting point is the 
N$^3$LO potential \cite{mach03}. 
From this potential we derive the corresponding $V_{\rm low-k}$ and use 
it directly in a HF calculation. 
The HF equations are then solved for $^{16}$O and $^{40}$Ca making use 
of a harmonic--oscillator basis.
We assume spherical symmetry, which implies that the HF SP states
$|\alpha\rangle$ have good orbital  and total angular momentum.
Therefore, they can be expanded in terms of oscillator wave functions
$|\mu \rangle$,

\begin{figure}[H]
\includegraphics[scale=1.0,angle=+90]{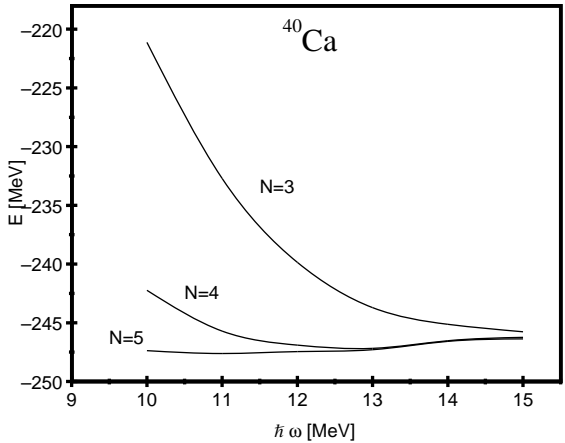}
\caption{Behavior of $E_{\rm HF}$ with $\hbar \omega$ and $N$ for 
$^{40}$Ca.}
\end{figure}

\begin{equation}
|\alpha \rangle = \sum_{\mu} C^{\alpha}_{\mu} | \mu \rangle~~,
\end{equation}

\noindent{where the sum is over the principal quantum number only.
The  expansion coefficients $C^{\alpha}_{\mu}$ are determined by solving
self--consistently the HF equations}

\begin{equation}
\sum_{\mu'} \langle \mu |t+U|\mu' \rangle C^{\alpha}_{
\mu'}=\epsilon_{\alpha} C^{\alpha}_{\mu}~~,
\end{equation}

\noindent{where $t$ is the kinetic energy and the HF potential $U$ is
defined as }

\begin{equation}
\langle \mu|U|\mu'\rangle = \sum_{\alpha_h}\langle \mu \alpha_h|V_{\rm
low-k}| \mu' \alpha_h\rangle~~,
\end{equation}

\noindent{with the index $\alpha_h$ referring to occupied states in the
ground--state  HF Slater determinant.}

Once  Eq. (7) has been solved, the  ground--state properties of the 
nucleus can be calculated. 
In particular, the total energy has  the well-known expression

\begin{equation}
E_{\rm HF}= \frac{1}{2} \sum_{\alpha_h} \left[ \langle \alpha_h
|t| \alpha_h \rangle + \epsilon_{\alpha_h} \right]~~.
\end{equation}

In our calculations the sum in the expansion (6) has been extended up 
to $N=5$ terms.
We have verified that this truncation is sufficient to ensure that the HF
results do not significantly depend on the variation of the oscillator 
constant $\hbar \omega$. 
This is illustrated in Figs. 1 and 2, where we show the behavior of the 
HF ground--state energy of $^{16}$O and $^{40}$Ca versus $\hbar \omega$ 
for different values of $N$. 
The results for $N=5$ are quite stable.
In our calculations the values of $\hbar \omega$ have been derived from
the expression $\hbar \omega= 45 A^{-1/3} -25 A^{-2/3}$ \cite{blomqvist68},
which reproduces the rms radii in an independent--particle approximation
with harmonic--oscillator wave functions.
This expression gives  $\hbar \omega= 14$ and 11 MeV for $^{16}$O and
$^{40}$Ca, respectively. 
In solving Eq. (7), the Coulomb potential has been added to $V_{\rm low-k}$
for protons.

\begin{figure}[H]
\includegraphics[scale=0.6,angle=-90]{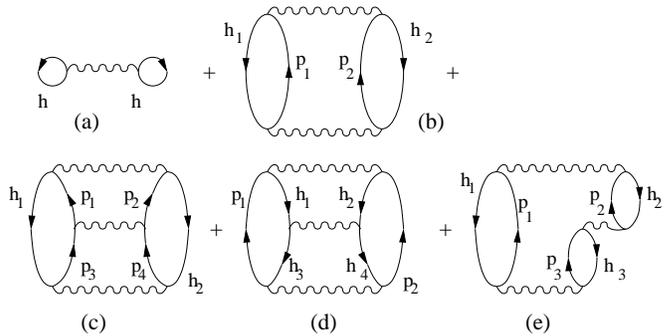}
\caption{First-, second-, and third-order diagrams in the Goldstone
expansion.}
\end{figure}

We use the HF basis to sum the Goldstone expansion including contributions
up to third order in $V_{\rm low-k}$.
In Fig. 3 we report the first--, second--,  and third--order diagrams
\cite{goldstone57}.
The intermediate states involved in the evaluation of these diagrams
are those obtained from the lowest 10 and 11 h.o. major shells for 
$^{16}$O and $^{40}$Ca, respectively.
Our results show convergence for these large spaces: for example, the
$^{16}$O binding energy per nucleon in second order approximation
is 7.03, 7.22, and 7.30 MeV when considering 9, 10, and 11 major shells, 
respectively.
Similar diagrams have been used to calculate the corrections to
the rms radius. 
As is well known, retaining only the first-order term in this expansion 
yields the HF results.

To conclude, it is worth stressing that using the Goldstone expansion in
terms of $V_{\rm low-k}$ one has to include also diagrams (b) and (c). 
This is not the case of the BG approach, where these diagrams are already 
contained in diagram (a) through the $G$ matrix.

\begin{table}[H]
\caption{Comparison of the calculated binding energy per nucleon 
(MeV/nucleon) and rms charge radius (fm) with the experimental data for 
$^{16}$O and $^{40}$Ca.}
\begin{ruledtabular}
\begin{tabular}{llcccc}
Nucleus &  & HF & HF+2nd  & HF+2nd+3rd  & Expt. \\
\colrule
 $^{16}$O & & & & & \\
 & $B/A$                 & 3.23 & 7.22 & 7.52 & 7.98 \\
 & $\langle r_c \rangle$ & 2.30 & 2.52 & 2.65 & $2.73 \pm 0.02$ \\
 $^{40}$Ca & & & & & \\
 & $B/A$                 & 6.19  & 9.10  & 9.19  & 8.55 \\
 & $\langle r_c \rangle$ & 2.610 & 3.302 & 3.444 & $3.485 \pm 0.003$ \\
\end{tabular}
\end{ruledtabular}
\end{table}

\section{Results}
In Table I, the calculated binding energy per nucleon and the rms charge
radius  for both $^{16}$O and $^{40}$Ca are compared with the experimental 
data \cite{audi93,devries87,nadjakov94}.
This Table contains the HF results as well as the values obtained
including second-- and third--order contributions. 
We see that the renormalization of the short range repulsion through 
$V_{\rm low-k}$ is sufficient to yield positive HF binding energies,
albeit too small as compared to the experimental values. 
We also see that the HF approximation significantly underestimates the 
rms radii.
This evidences the role  of higher--order contributions in the Goldstone 
expansion to account for correlations beyond the mean field. 
The binding energies and radii calculated including diagrams up to third 
order are very satisfactory.
In fact, the binding energies gain about 4 and 3 MeV for $^{16}$O and
$^{40}$Ca, respectively, while the radii increase by about 0.4 and 0.8 fm 
coming quite close to the experimental values.

\begin{table}[H]
\caption{Calculated and experimental single-particle energies (MeV) for
$^{16}$O. Experimental data are taken from \cite{audi93,nndc,lodhi74}.}
\begin{ruledtabular}
\begin{tabular}{cccccc}
 ~ & ~ & ~ & $^{16}$O  & ~ & ~ \\
  ~ & \multicolumn{2}{c}{Neutron} & ~ & \multicolumn{2}{c}{Proton} \\  
 Orbital & Calc. & Expt. & ~ & Calc. & Expt. \\
\colrule
 $ s_{1/2}$   & $-53.786$ & $-47$     & ~ & $-48.606$ & $-44 \pm 7$\\
 $ p_{3/2}$   & $-23.225$ & $-21.839$ & ~ & $-19.264$ & $-18.451$  \\
 $ p_{1/2}$   & $-15.649$ & $-15.663$ & ~ & $-11.841$ & $-12.127$  \\
 $ d_{5/2}$   & $ -0.056$ & $ -4.144$ & ~ & $  3.564$ & $ -0.601$  \\
 $ s_{1/2}$   & $ -0.481$ & $ -3.273$ & ~ & $  2.651$ & $ -0.106$  \\
 $ d_{3/2}$   & $  5.814$ & $  0.941$ & ~ & $  8.713$ & $  4.399$  \\
\end{tabular}
\end{ruledtabular}
\end{table}

A discussion of the convergence properties of the perturbative series
is now in order.
To this end, the HF potential energy, the second-- , and third--order
corrections have to be compared.
In $^{16}$O the HF potential energy per nucleon is $V^{(1)}=-23.5$ MeV,
which is obtained by subtracting from the total HF energy the contribution 
of the kinetic term. 
Thus for the ratio of the second-- to first--order term we obtain 
$V^{(2)}/V^{(1)}=0.17$, while the ratio $V^{(3)}/V^{(2)}$ is 0.08. 
Similarly, for $^{40}$Ca we have $V^{(1)}=-33.7$ MeV, 
$V^{(2)}/V^{(1)}=0.09$, and $V^{(3)}/V^{(2)}=0.03$. 
On these grounds, we may conclude that the convergence of the series is 
fairly rapid and that higher--order contributions are negligible.

For the sake of completeness, in Tables II and III we report the HF 
single-hole energies as well as the energies of the low-lying particle
states of $^{16}$O and $^{40}$Ca. 
In Tables IV and V the calculated occupation probabilities for states 
up to the Fermi level are reported.

\begin{table}[H]
\caption{Calculated and experimental single-particle energies (MeV) for
$^{40}$Ca. Experimental data are taken from \cite{audi93,nndc,lodhi74}.}
\begin{ruledtabular}
\begin{tabular}{cccccc}
 ~ & ~ & ~ & $^{40}$Ca  & ~ & ~ \\
  ~ & \multicolumn{2}{c}{Neutron} & ~ & \multicolumn{2}{c}{Proton} \\  
 Orbital & Calc. & Expt. & ~ & Calc. & Expt. \\
\colrule
 $ s_{1/2}$   & $-97.944$ &   ~     & ~ & $-87.501$ & $\begin{array}{c}
-49.1 \pm 12 \\ -77 \pm 14 \\ \end{array}$ \\
 $ p_{3/2}$   & $-63.760$ &   ~      & ~ & $-53.934$ & $-33.3 \pm 6.5$ \\
 $ p_{1/2}$   & $-54.959$ &   ~      & ~ & $-45.269$ & $-32 \pm 4$ \\
 $ d_{5/2}$   & $-33.018$ & $-21.30$ & ~ & $-23.749$ & $ \begin{array}{c} 
-14.9 \pm 2.5 \\ -13.8 \pm 7.5 \\ \end{array}$ \\
 $ s_{1/2}$   & $-27.406$ & $-18.104$ & ~ & $-18.238$ & $-10.850$ \\
 $ d_{3/2}$   & $-19.595$ & $-15.641$ & ~ & $-10.663$ & $ -8.328$ \\
 $ f_{7/2}$   & $ -6.579$ & $-8.363$ & ~ & $  2.047$ &  $ -1.085$ \\
 $ p_{3/2}$   & $ -4.325$ & $-6.420$ & ~ & $  3.293$ &  $  0.631$ \\
 $ p_{1/2}$   & $ -0.973$ & $~~$     & ~ & $  5.865$ &  $~~$      \\
 $ f_{5/2}$   & $  5.852$ & $~~$     & ~ & $ 12.484$ &  $~~$      \\
\end{tabular}
\end{ruledtabular}
\end{table}

\section{Concluding remarks}

In this work, starting from the chiral N$^3$LO potential \cite{mach03}, 
we have performed calculations for the ground-state properties
of the doubly closed nuclei $^{16}$O and $^{40}$Ca making use of
the Goldstone expansion.
This has been done within the framework of a new approach 
\cite{bogner01,bogner02} to the renormalization of the short-range 
repulsion of realistic $NN$ potentials, wherein a low-momentum potential 
$V_{low-k}$ is constructed which preserves the low-energy physics of the 
original potential.
We consider a main achievement of our study to have shown that $V_{\rm
low-k}$ is suitable for being used directly in the Goldstone expansion.
Namely, unlike the traditional BHF approach, there is no need to first
calculate the $G$ matrix. 
We have seen that taking into account higher-order contributions 
(essentially the second-order terms) of  $V_{\rm low-k}$ in the 
Goldstone expansion yields very good results for the binding energy and
charge radius of $^{16}$O and $^{40}$Ca.

\begin{table}[H]
\caption{Calculated occupation probabilities for $^{16}$O.}
\begin{ruledtabular}
\begin{tabular}{cccc}
 Orbital & Proton & ~ & Neutron \\
\colrule
 $ s_{1/2}$   & $0.881$ & ~ & $0.881$ \\
 $ p_{3/2}$   & $0.822$ & ~ & $0.818$ \\
 $ p_{1/2}$   & $0.767$ & ~ & $0.760$ \\
\end{tabular}
\end{ruledtabular}
\end{table}

In this context, it is worth emphasizing that the idea of bypassing the
$G$-matrix approach to nuclear structure calculations is not new
\cite{tabakin64,elliott68}. 
From this viewpoint, the present study comes close in spirit to the early 
work carried out at the MIT \cite{kerman66,kerman67,bassichis67} in the 
mid 1960's. 
There the Tabakin's separable nonlocal $NN$ potential \cite{tabakin64} 
was used in HF calculations and then the second-order correction to the 
binding energy was evaluated, obtaining quite satisfactory results. 
It is indeed very gratifying that our results based on the use of the 
$V_{\rm low-k}$ derived from the modern N$^3$LO potential come 
significantly closer to the experimental data.
It should be noted that our results are quite good also when compared to
those of recent BHF calculations \cite{gad02}, where different modern 
$NN$ potentials have been used and long-range cortrelations have been 
considered within the framework of the Green function approach.

\begin{table}[H]
\caption{Calculated occupation probabilities for $^{40}$Ca.}
\begin{ruledtabular}
\begin{tabular}{cccc}
 Orbital & Proton & ~ & Neutron \\
\colrule
 $ s_{1/2}$   & $0.945$ & ~ & $0.947$ \\
 $ p_{3/2}$   & $0.931$ & ~ & $0.932$ \\
 $ p_{1/2}$   & $0.921$ & ~ & $0.922$ \\
 $ d_{5/2}$   & $0.885$ & ~ & $0.884$ \\
 $ s_{1/2}$   & $0.858$ & ~ & $0.855$ \\
 $ d_{3/2}$   & $0.811$ & ~ & $0.807$ \\
\end{tabular}
\end{ruledtabular}
\end{table}

In summary, we may conclude that the results of the present study, 
together with those of our recent shell-model calculations, show that 
the $V_{\rm low-k}$ approach provides a simple and reliable way of 
``smoothing out" the repulsive core contained in the modern $NN$ 
potentials before using them in microscopic nuclear structure 
calculations.

\begin{acknowledgments}
This work was supported in part by the Italian Ministero
dell'Istruzione, dell'Universit\`a e della Ricerca  (MIUR) and by the
U.S. DOE Grant No.~DE-FG02-88ER40388. We would like to thank R. Machleidt
for providing us with the matrix elements of the N$^3$LO potential.
\end{acknowledgments}


\begin{references}

\bibitem{cdbonn} R. Machleidt, Phys. Rev. C {\bf 63}, 024001 (2001).

\bibitem{stoks94} V. G. J. Stoks, R. A. M. Klomp, C. P. F. Terheggen,
and J. J. de Swart, Phys. Rev. C {\bf 49}, 2950 (1994).

\bibitem{wiringa95} R. B. Wiringa, V. G. J. Stoks, and R. Schiavilla,
Phys. Rev. C {\bf 51}, 38 (1995).

\bibitem{mach03} D. R. Entem and  R. Machleidt, nucl-th/0304018.

\bibitem{Day67} See, for instance, B. D. Day, Rev. Mod. Phys. {\bf 39}, 
719 (1967).

\bibitem{towner} I. S. Towner, {\em A Shell Model Description of Light
Nuclei} (Clarendon Press, Oxford, 1977).

\bibitem{sch91} K. W. Schmid, H. M\"uther, and R. Machleidt, Nucl. Phys. 
A {\bf 530}, 14 (1991), and references therein.

\bibitem{gad02} Kh. Gad and H. M\"uther, Phys. Rev. C {\bf 66}, 044301
(2002).

\bibitem{fabrocini00} A. Fabrocini, F. Arias de Saavedra, and G. C\'o,
Phys. Rev. C {\bf 61}, 044302 (2000), and references therein.

\bibitem{heisenberg99} Jochen H. Heisenberg and Bogdan Mihaila,
Phys. Rev. C {\bf 59}, 1440 (1999), and references therein.

\bibitem{muther00} H. M\"uther and A. Polls,
Prog. Part. Nucl. Phys. {\bf 45}, 243 (2000).

\bibitem{suzuki94} K. Suzuki and R. Okamoto, Prog. Theor. Phys.
{\bf 92}, 1045 (1994).

\bibitem{bogner01} S. Bogner, T.T.S. Kuo and L. Coraggio, Nucl. Phys.
A {\bf 684}, 432c (2001).

\bibitem{bogner02} Scott Bogner, T. T. S. Kuo, L. Coraggio,
A. Covello, and N. Itaco, Phys. Rev. C {\bf 65}, 051301(R) (2002).

\bibitem{covello02} A. Covello, L. Coraggio, A. Gargano, N. Itaco, and
T.T. S. Kuo, in {\it Challenges of Nuclear Structure},
Proceedings of the Seventh International Spring Seminar on Nuclear Physics,
Maiori, Italy, 2001, edited by A. Covello (World Scientific, Singapore,
2002),
p. 139.

\bibitem{coraggio02} L. Coraggio, A. Covello, A. Gargano, N. Itaco,
T.T. S. Kuo, D. R. Entem, and R. Machleidt, Phys. Rev. C {\bf 66},
021303(R) (2002).

\bibitem{coraggio02a} L. Coraggio, A. Covello, A. Gargano, N. Itaco, and
T.T. S. Kuo, Phys. Rev. C {\bf 66},
064311 (2002).

\bibitem{mach02}D. R. Entem and R. Machleidt, Phys. Lett. B {\bf 524},
93 (2002);
D. R. Entem, R. Machleidt, and H. Witala, Phys. Rev.
C {\bf 65}, 064005 (2002).

\bibitem{krenc80} E. M. Krenciglowa and T. T. S. Kuo, Nucl. Phys. A
{\bf 342}, 454 (1980).

\bibitem{andre96} F. Andreozzi, Phys. Rev. C {\bf 54}, 684 (1996).

\bibitem{blomqvist68} J. Blomqvist and A. Molinari, Nucl. Phys. A {\bf
106}, 545 (1968).

\bibitem{goldstone57} J. Goldstone, Proc. Roy. Soc. (London)
A {\bf 239}, 267 (1957).

\bibitem{audi93} G. Audi and A. H. Wapstra, Nucl. Phys. A {\bf 565}, 1
(1993).

\bibitem{devries87} H. de Vries, C. W. de Jager, and C. de Vries,
At. Data Nucl. Data Tables {\bf 36}, 495 (1987).

\bibitem{nadjakov94} E. G. Nadjakov, K. P. Marinova, and Yu. P. Gangrsky,
At. Data Nucl. Data Tables {\bf 56}, 133 (1994).

\bibitem{nndc} Data extracted using the NNDC On-Line Data Service from
the ENSDF database, files revised as of December 5, 2001, M.R. Bhat, {\em
Evaluated Nuclear Structure Data File (ENSDF), Nuclear Data for
Science and Technology}, edited by S. M. Quaim (Springer-Verlag,
Berlin, Germany, 1992), p. 817.

\bibitem{lodhi74} M. A. K. Lodhi and B. T. Waak, Phys. Rev. Lett. {\bf 33}, 
431 (1974).

\bibitem{tabakin64} F. Tabakin, Ann. Phys. (N.Y.) {\bf 30}, 51 (1964); Phys.
Rev. {\bf 174}, 1208 (1968).

\bibitem{elliott68} J. P. Elliott, A. D. Jackson, H. A.
Mavromatis, E. A. Sanderson, and B. Singh,
Nucl. Phys. A {\bf 121}, 241 (1968).

\bibitem{kerman66} A. K. Kerman, J. P. Svenne, and F. M. H. Villars,
Phys. Rev. {\bf 147}, 710 (1966).

\bibitem{kerman67} A. K. Kerman and M. K. Pal, Phys. Rev. {\bf 162},
970 (1967).

\bibitem{bassichis67} W. H. Bassichis, A. K. Kerman, and J. P. Svenne,
Phys. Rev. {\bf 160} 746 (1967).

\end{references}
\end{document}